\documentclass[aps,pra,amsmath,amssymb,onecolumn,superscriptaddress,showpacs,floatfix,longbibliography]{revtex4-1}
\usepackage[english]{babel}
\usepackage[utf8]{inputenc}
\usepackage[colorinlistoftodos, color=green!40, prependcaption]{todonotes}
\usepackage{amsthm}
\usepackage{mathtools}
\usepackage{physics}
\usepackage{xcolor}
\usepackage{graphicx}
\usepackage[left=23mm,right=13mm,top=35mm,columnsep=15pt]{geometry} 
\usepackage{adjustbox}
\usepackage{placeins}
\usepackage[T1]{fontenc}
\usepackage{lipsum}
\usepackage{csquotes}
\usepackage[pdftex, pdftitle={Article}, pdfauthor={Author}]{hyperref} 
\begin{document}
\title{Dephasing-assisted entanglement in a system of strongly coupled qubits}

\author{I. V. Vovcenko}
    \affiliation{Moscow Institute of Physics and Technology, 141700, Moscow region, Russia}
    \affiliation{Institute for Theoretical and Applied Electromagnetics, 125412, Moscow, Russia}

\author{V. Yu. Shishkov}
    \affiliation{Moscow Institute of Physics and Technology, 141700, Moscow region, Russia}
    \affiliation{Institute for Theoretical and Applied Electromagnetics, 125412, Moscow, Russia}
    \affiliation{Dukhov Research Institute of Automatics, 127055, Moscow, Russia}
    
\author{E. S. Andrianov}
    \email{andrianov.es@mipt.ru}
    \affiliation{Moscow Institute of Physics and Technology, 141700, Moscow region, Russia}
    \affiliation{Institute for Theoretical and Applied Electromagnetics, 125412, Moscow, Russia}
    \affiliation{Dukhov Research Institute of Automatics, 127055, Moscow, Russia}
    
\date{\today} 

\maketitle

\section*{abstract}
Creation of entangled states of quantum systems with low decoherence rates is a cornerstone in practical implementation of quantum computations. Processes of separate dephasing in each qubit in experimentally feasible systems is commonly accepted to destroy entanglement. In this work, we consider a system of two strongly coupled qubits that interact with dephasing reservoirs. We demonstrate that interaction with dephasing reservoirs can contribute to the formation of a long-lived mixed entangled state with nonzero concurrence. The weight of the subradiant state in this mixed state tends toward unity if the dephasing rate is much larger than the radiative rate and less than the coupling constant between qubits. The lifetime of this state is proportional to the exponent of the ratio of the coupling constant to environmental temperature and can be, by orders of magnitude, larger than the system's characteristic dephasing and dissipation times. Therefore, high dephasing, along with strong coupling, contributes to the creation of an entangled state with a long lifetime. This result paves the way for creation of long-lived entangled states.

\section{Introduction}
Entangled states of two or more qubits are the building blocks of quantum computers and elements of quantum communication systems \cite{1}. The concept of entanglement was first proposed for the pure state, which can be described in terms of wave function \cite{2,3,4,5}. However, all real systems inevitably interact with the environment. This leads to a state of the system ceases to be pure and becomes a mixed state, which can be described only in terms of a density matrix. The concept of entanglement can be extended to mixed states, as well \cite{6}. The system-environment interaction leads to processes of dissipation and dephasing or decoherence. The first leads to change in both energy and coherence, while the second process does not change the energy and results in destruction of coherence. This manifests as dissipation of only the non-diagonal density matrix  elements. In real systems, the dephasing rate is up to five orders of magnitude higher than the dissipation rate \cite{7,8} and it is responsible for the fast destruction of the entanglement.

One basic element for entanglement creation is an ensemble of two two-level systems (TLSs) interacting with the environment and, possibly, each other \cite{9,10,11,12,13}. The problem of entanglement creation in such a system has been investigated in many works \cite{14,15,16,17,18,19,20}, and several opportunities to create and conserve entanglement exist. One can enhance the interaction between TLSs in such a way that the ground state of the system becomes entangled \cite{17,20} or use coherent external drive to move the system into the entangled state \cite{21,cecoi2018entanglement}. The use of common reservoirs for two qubits is another way to create entanglement \cite{14,16,18,22}. 

Free space modes of electromagnetic field are an example of common reservoirs for dipole moments of atoms or molecules. When they occupy subwavelength volume, an entangled subradiant state, which is antisymmetric with respect to mutual permutation, is protected from interaction with a common radiative reservoir. For this reason, the creation of a subradiant state is particularly interesting. There are quantum algorithms and quantum memory schemes based on manipulation with such a state \cite{23,24,25,26,27}. Additionally, one can use interfaces, e.g., waveguides \cite{28}, plasmonic waveguides \cite{29,30} or chiral waveguides \cite{31}, to move the system, more effectively, towards the subradiant state. 

Thus, a common reservoir can contribute to the formation of entanglement. This concerns the dephasing reservoir as well. If one considers dephasing reservoirs common for two subsystems, long-lived entanglement is possible \cite{19}. At the same time, dephasing reservoirs usually are separate reservoirs because they are associated with nuclear vibration degrees of freedom for each molecule or a quantum dot \cite{32,33}. Thus, such type of reservoirs are expected to result in the entanglement destruction. 

However, when the system contains interacting subsystems, relaxation of one system inevitably leads to relaxation of the others. The last statement can be elucidated in the Born-Markov approximation by means of a master equation in the Lindblad form. As the Lindblad approach gives an answer consistent with thermodynamic laws, using eigenstates of the whole system is necessary \cite{34,35}. In the case of well-separated eigenstates, the Lindblad superoperator describes the transitions between eigenstates. Because the eigenstate of the interacting subsystem is a superposition of isolated subsystem eigenstates, mentioned Lindblad superoperators will result in a cross-relaxation process when relaxation of one subsystem leads to relaxation of the others \cite{34,shishkov2020perturbation}. Recently, construction of the correct master equation for the system of coupled TLSs, interacting with different types of reservoirs, was actively investigated \cite{36,37,38}. In view of the above, revising the problem of two-qubit entanglement, in case of strong coupling, using recent progress in physics of open quantum systems is important \cite{38}.

In this paper, we consider the system of two strongly coupled qubits each of which interacts with its own dephasing reservoir. We show that dephasing not only results in the relaxation of non-diagonal terms of density matrix, but also leads to the transition between super- and sub-radiant states. If coupling constant between qubits is much larger than dephasing rate then there is exist long-lived mixed entangled state. Its lifetime is proportional to the exponent of ratio of coupling constant to environmental temperature. If coupling constant between qubits is much larger than the temperature of dephasing reservoir, the lifetime of the mentioned state is much larger than the characteristic dephasing and dissipation times of the system. It is demonstrated that the weight of entangled subradiant state in this mixed state tends to unity, if dephasing rate is much higher than the radiative rate, i.e., high dephasing contributes to entanglement. As a result, concurrence of this state is greater than zero which is a criterion of a state to be entangled. The obtained results open the way to use dephasing as a source for the creation of long-lived entanglement.

\section{Formation of long-lived entanglement by separate dephasing resevoirs in the system of strongly-coupled qubits}

We consider the system of two strongly coupled TLSs with transition frequencies ${\omega _1} \approx {\omega _2}$ lying in the optical range and with the distance between TLSs $r$ that is much smaller than the optical wavelength, i.e. $r \ll \lambda_{1,2} = 2 \pi c / \omega_{1,2}$. We suppose that both TLSs interact with a common radiative reservoir, and each TLS interacts with its own dephasing reservoir. Each dephasing reservoir is associated with vibrational degrees of freedom (e.g. phonons ~\cite{33}) of qubits. Because we are interested in the case of nearby qubit frequencies, $\left| {{\omega _1} - {\omega _2}} \right| \ll {\omega _{1,2}}$ we will use the rotating wave approximation \cite{39}. Thus, the Hamiltonian of the system takes the form \cite{32,33}:
\begin{equation}
{\hat H_{\rm{S}}} = \hbar {\omega _1}\hat \sigma _1^\dag {\hat \sigma _1} + \hbar {\omega _2}\hat \sigma _2^\dag {\hat \sigma _2} + \hbar \Omega \left( {\hat \sigma _1^\dag {{\hat \sigma }_2} + \hat \sigma _2^\dag {{\hat \sigma }_1}} \right).
\label{1}
\end{equation}
Here, ${\hat \sigma _{1,2}}$ are lowering operators for the first and second TLSs, respectively. The last term in Eq. (\ref{1}) describes the dipole-dipole interaction in the rotating wave approximation when dipoles occupy subwavelength volume \cite{agarwal1974quantum}, $\Omega  = \left( {{{\bf{d}}_1}{{\bf{d}}_2} - 3\left( {{{\bf{d}}_1}{\bf{n}}} \right)\left( {{{\bf{d}}_2}{\bf{n}}} \right)} \right)/\hbar {r^3}$ is the constant of interaction, ${{\bf{d}}_{1,2}}$  are dipole transition matrix elements of the first and second TLSs, and ${\bf{n}}$ is the unit vector from one TLS to another.
First, we consider the case of zero detuning, ${\omega _1} = {\omega _2} = \omega$, and ${\Omega} > 0$. As such, the system eigenstates are the excited state of two TLSs, $\left| {ee} \right\rangle $, the superradiant symmetric state $\left| s \right\rangle=\left( {\left| {eg} \right\rangle  + \left| {ge} \right\rangle } \right)/\sqrt{2}$, the subradiant antisymmetric state $\left| as \right\rangle=\left( {\left| {eg} \right\rangle  - \left| {ge} \right\rangle } \right)/\sqrt 2 $, and the ground state $\left| {gg} \right\rangle $,  with the eigenfrequencies $2\omega$, $\omega  + \Omega $, $\omega  - \Omega $, and $0$, respectively.

The master equation for the system density matrix can be obtained by the Born-Markov approximation via standard procedure \cite{32,35,41}, assuming that the reservoirs are in thermal equilibrium. As a result, we obtain the following Lindblad equation (see Appendix):
\begin{equation}
\begin{array}{l}
{{\dot{\hat{\rho}} }_{\rm{S}}}(t) =  - i{\hbar ^{ - 1}}[{{\hat H}_{\rm{S}}},{{\hat \rho }_{\rm{S}}}(t)] +
  \sum\limits_{{\rm{k}} = 1,2} {\sum\limits_{{\rm{j}} = 1,2,3} {\frac{{{\gamma _{{\rm{k}}{\rm{,dp}}}}\left( {2\Omega {\theta _{\rm{j}}}} \right)}}{2} \mathcal{L}_{\rm{dp,kj}} \left[\hat \rho\right]} }  + \\
 + \sum\limits_{{\rm{k}} = 1,2} {\frac{{{\gamma _{{\rm{rad}}}}\left( { - \left( {\omega  + {{( - 1)}^{{\rm{k}} - 1}}\Omega } \right)} \right)}}{2} \mathcal{L}_{\rm{rad,k}} \left[\hat \rho\right]} 
 + \sum\limits_{{\rm{k}} = 1,2} {\frac{{{\gamma _{{\rm{rad}}}}\left( {\left( {\omega  + {{( - 1)}^{{\rm{k}} - 1}}\Omega } \right)} \right)}}{2} \mathcal{L}_{\rm{rad,k}} \left[\hat \rho\right]} {\kern 1pt} ,
\end{array}
\label{2}
\end{equation}
where Lindblad superoperators have the standard form, ${{\cal L}_i}\left[ {\hat \rho } \right] = 2{{\hat L}_i}{{\hat \rho }_{\rm{S}}}(t)\hat L_i^\dag  - {{\hat \rho }_{\rm{S}}}(t)\hat L_i^\dag {{\hat L}_i} - \hat L_i^\dag {{\hat L}_i}{{\hat \rho }_{\rm{S}}}(t)$, and ${\cal L}_i^\dag \left[ {\hat \rho } \right] = 2\hat L_i^\dag {{\hat \rho }_{\rm{S}}}(t){{\hat L}_i} - {{\hat \rho }_{\rm{S}}}(t){{\hat L}_i}\hat L_i^\dag  - {{\hat L}_i}\hat L_i^\dag {{\hat \rho }_{\rm{S}}}(t)$. Operators $\hat L_i$ are found to be 
\begin{equation}
\begin{array}{l}
{{\hat L}_{{\rm{rad}}{\rm{,1}}}} = {{\hat \sigma }_1} + {{\hat \sigma }_2} - \hat \sigma _1^\dag {{\hat \sigma }_1}{{\hat \sigma }_2} - {{\hat \sigma }_1}\hat \sigma _2^\dag {{\hat \sigma }_2},\ 
{{\hat L}_{{\rm{rad}}{\rm{,2}}}} = \hat \sigma _1^\dag {{\hat \sigma }_1}{{\hat \sigma }_2} + {{\hat \sigma }_1}\hat \sigma _2^\dag {{\hat \sigma }_2},\\
{{\hat L}_{{\rm{dp}}{\rm{,11}}}} = {{\hat L}_{{\rm{dp}}{\rm{,21}}}} = \hat \sigma _1^\dag {{\hat \sigma }_1}/2 + \hat \sigma _2^\dag {{\hat \sigma }_2}/2,\ 
{{\hat L}_{{\rm{dp}}{\rm{,12}}}} =  - {{\hat L}_{{\rm{dp}}{\rm{,22}}}} = \left( {\hat \sigma _1^\dag  + \hat \sigma _2^\dag } \right)\left( {{{\hat \sigma }_1} - {{\hat \sigma }_2}} \right)/4,\\
{{\hat L}_{{\rm{dp}}{\rm{,13}}}} =  - \,{{\hat L}_{{\rm{dp}}{\rm{,23}}}} = \hat L_{{\rm{dp}}{\rm{,12}}}^\dag,
\end{array}
\label{3}
\end{equation}
where parameter ${\theta _{\rm{j}}} = 0,\,1,\,-1$ for ${\rm{j}} = 1,\,2,\,3$. The dissipation constants, $\gamma_i$, are determined according to  ${\gamma _i}\left( \omega  \right) = \int\limits_{ - \infty }^\infty  {d\tau \exp \left( { - i\omega \tau } \right)\left\langle {{\hat{\tilde{R}}_i}(t + \tau ){\hat{\tilde {R}}_i}(t)} \right\rangle } $ \cite{34,35}, which, for a reservoir in thermal equilibrium at temperature $T$, satisfies the Kubo-Martin-Shwinger (KMS) condition ${\gamma _i}\left( \omega  \right) = \exp \left( { - \hbar \omega /k{T_i}} \right){\gamma _i}\left( { - \omega } \right)$.

The main consequence of usage of correct Lindblad operators (\ref{3}) is that dephasing not only results in the relaxation of non-diagonal density matrix terms (in the eigenstate basis) but also leads to the transition between super- and subradiant states. Indeed, operators ${\hat L_{\rm{dp},11}},\;{\hat L_{\rm{dp},21}}$  describe dephasing of both TLSs while operators ${\hat L_{\rm{dp},12}},\;{\hat L_{\rm{dp},22}}$ and ${\hat L_{\rm{dp},13}},\;{\hat L_{\rm{dp},23}}$ describe forward and backward energy flows between the system and dephasing reservoirs, respectively. Notably, these operators have nonzero matrix elements between the system eigenstates $\left| s \right\rangle$  and $\left| as \right\rangle$. This means that dephasing reservoirs result in transitions between super- and subradiant states. The reason for the energy flow between the system and the dephasing reservoir is interaction between the TLSs and expansion of the system reservoir interaction operators over eigenstates of interacting TLSs. These terms do not appear in the local approach, in which Lindblad superoperators are assumed as in the case of non-interacting TLSs \cite{34,shishkov2020perturbation}. Note that in \cite{cecoi2018entanglement} it has been mentioned that the dephasing reservoirs in the case of coupled qubits result in transition between super- and subradiant states. However, in \cite{cecoi2018entanglement} it has been considered stationary entanglement which is created by the external coherent drive. In contrast, here we are interested in the temporal dynamics of the entanglement between two qubits without any external coherent source.  

Operator ${\hat L_{\rm{rad},1}}$  defines the influence of the radiative reservoir and consist of two parts. The first part represents relaxation from the state  $\left| {ee} \right\rangle $ to the symmetric state $\left| s \right\rangle$  and from the state  $\left| s \right\rangle$ to the ground state $\left| {gg} \right\rangle $. The second part is energy flow from one TLS to another. Operator ${\hat L_{\rm{rad},2}}$  describes relaxation of nondiagonal density matrix terms. 

Note that the matrix elements between the states $\left| as \right\rangle$  and  $\left| {gg} \right\rangle$ of the Lindblad superoperators, arising from both dephasing and radiative reservoirs, are equal to zero. Thus, neither radiative nor dephasing reservoirs lead to transition from the subradiant state to the ground state. The only possibility of the system to leave the state  $\left| as \right\rangle$ is a temperature induced transition to the state  $\left| s \right\rangle$ which has a higher eigenfrequency. According to the KMS condition, the rate of this transition is smaller than the rate of transition from the state  $\left| s \right\rangle$ to the state $\left| as \right\rangle$  by the factor  $\exp \left( { - 2\hbar {\Omega}/kT} \right)$. From these qualitative arguments, it follows that if the system reaches the state  $\left| as \right\rangle$, then the time during which the system remains in this state is larger than the characteristic relaxation times by the factor $\exp \left( {2\hbar {\Omega}/kT} \right)$. Because this state is entangled, one can expect that entanglement between two TLSs will also be conserved during the time that is larger than the characteristic relaxation times by the factor $\exp \left( {2\hbar {\Omega}/kT} \right)$.

\begin{figure*}[ht]
\begin{minipage}[h]{0.49\linewidth}
\center{\includegraphics[width=0.72\linewidth]{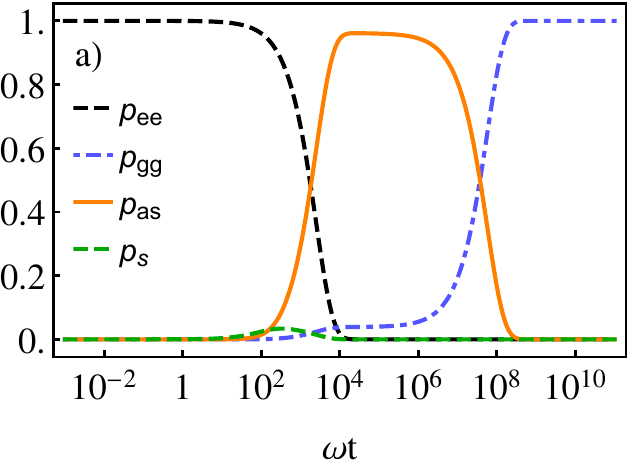}}
\end{minipage}
\hfill
\begin{minipage}[h]{0.49\linewidth}
\center{\includegraphics[width=0.72\linewidth]{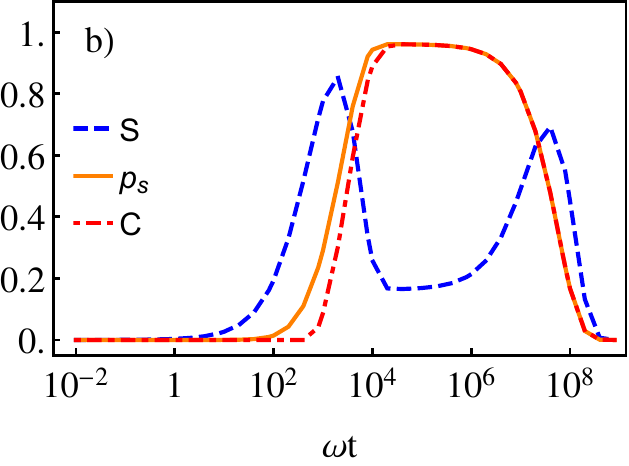} }
\end{minipage}
\caption{ The dependencies of density matrix elements $p_{ee}$ (black dashed line), $p_{gg}$ (blue dash-dotted line), $p_{as}$ (orange solid line), $p_{s}$ (green dashed line) (a), and entropy (blue dashed line), concurrence (red dash-dotted line), antisymmetric state occupancy (orange solid line) (b) on time when the system interacts with two separate dephasing reservoirs and common radiative reservoir.}
\label{TwoRes}
\end{figure*}

Numerical simulations and analytical evaluations confirm these qualitative arguments. 
Results of the simulation of master equation (\ref{2}), with the initial condition $\hat \rho_S (0) = | ee \rangle \langle ee |$, are shown in the Fig. \ref{TwoRes}. Experimentally feasible parameters for molecules and quantum dots are  ${\gamma _{\rm{dp}}} \simeq {10^{11}}{\ \rm{s}^{ - 1}}$, ${\gamma _{\rm{rad}}} \simeq {10^{8}} - {10^{9}}{\ \rm{s}^{ - 1}}$. The value of the Rabi constant depends on the dipole moment matrix element. We are interested in the case of strong coupling. Thus, we consider the case of quantum dots with a large dipole moment $ \simeq 50{\ \rm{D}}$ \cite{7,8} and $ \simeq 10{\ \rm{nm}}$ distance between them, such that the corresponding Rabi constant is ${\Omega} \simeq {2 \cdot 10^{12}}{\ \rm{s}^{ - 1}}$. In numerical simulations, we use the value $\simeq {\rm{0.01 \ eV}}$  as a dimensionless unit of frequency such that the other parameters are ${\gamma _{\rm{dp}1,2}}\left( { - 2{\Omega}} \right) =\gamma _{\rm{dp}}= {2\cdot10^{ - 2}}$, ${\gamma _{\rm{rad}}}\left( { - \omega  \pm \Omega } \right) =\gamma _{\rm{rad}}= {2\cdot10^{ - 4}}$, ${T_{\rm{dp}}} = {T_{\rm{rad}}} = 2 \cdot {10^{ - 2}}$ ($\simeq 5{\rm K}$), and $\Omega  = {10^{ - 1}}$. 

In Fig. \ref{TwoRes}a, the dependencies of the non-zero matrix elements of the density matrix are shown, namely, the probabilities ${p_{ee}}$, ${p_{as}}$, ${p_{s}}$, and  ${p_{gg}}$ of the system to be in the states $\left| {ee} \right\rangle $, $\left| as \right\rangle$, $\left| s \right\rangle$, and $\left| {gg} \right\rangle $, respectively. There are two stages in the system dynamics. During the first stage ($t \le {10^5}$), the system starting from the state $\left| {ee} \right\rangle $  relaxes and primarily occupies the antisymmetric subradiant state. At the second stage, ($t \ge {10^5}$), the system leaves this state and relaxes to the ground state of the system, $\left| {gg} \right\rangle $. We calculated the entropy, $S =  - {\rm{Tr}}\left( {\hat{\rho}_{\rm{S}} \log \hat{\rho}_{\rm{S}}} \right)$, and concurrence, $C = \max \{ 0,\sqrt {{\lambda _1}}  - \sqrt {{\lambda _2}}  - \sqrt {{\lambda _3}}  - \sqrt {{\lambda _4}} \}$,    where  ${\lambda _i}$ are eigenvalues of the matrix $\hat \rho_{\rm{S}} \left[ {\left( {{{\hat \sigma }_y} \otimes {{\hat \sigma }_y}} \right)\hat \rho_{\rm{S}}^*\left( {{{\hat \sigma }_y} \otimes {{\hat \sigma }_y}} \right)} \right]$ \cite{4}. The evolution of entropy, concurrence and occupancy of the antisymmetric state are presented in Fig. \ref{TwoRes}b. 

At the first stage, entropy of the system grows (Fig. \ref{TwoRes}b, blue dashed line) because the system transits from the state  $\left| {ee} \right\rangle $ to the subspace formed by states $\left| {eg} \right\rangle $  and $\left| {ge} \right\rangle $. After the first stage finishes, the system primarily occupies the antisymmetric state (Fig. \ref{TwoRes}b, orange solid line) and stays in this state for a time which is much larger than all relaxation times in the system. As a consequence, the entropy at the second stage decreases while concurrence grows from zero to 0.95 (Fig. \ref{TwoRes}b, red dash-dotted line). This indicates that the system is in the entangled state. Finally, the system relaxes to the ground state which is also accompanied by the temporary grows of entropy. 

As has been mentioned in the introduction, in general, separate reservoirs destruct entanglement. In our case, the main contribution to the formation of entanglement is made by separated dephasing reservoirs. We illustrate this fact with two examples \ref{A} and \ref{B} described below. 
Additionally, we examine the system with help of local approach in the third example \ref{C} to show that the local approach does not reveal the entanglement.

\subsection{Long-lived entanglement in the case of one dephasing reservoir}\label{A}

In the first example, we turn off the dephasing reservoirs associated with the first TLS, while the dephasing reservoir associated with the second TLS is not changed (i.e., $\gamma_{\rm{1,dp}}=0$ in Eq. (\ref{2})). In other words, we break the symmetry of interaction between the TLSs and dephasing reservoirs. 

\begin{figure*}[ht]
\begin{minipage}[h]{0.49\linewidth}
\center{\includegraphics[width=0.72\linewidth]{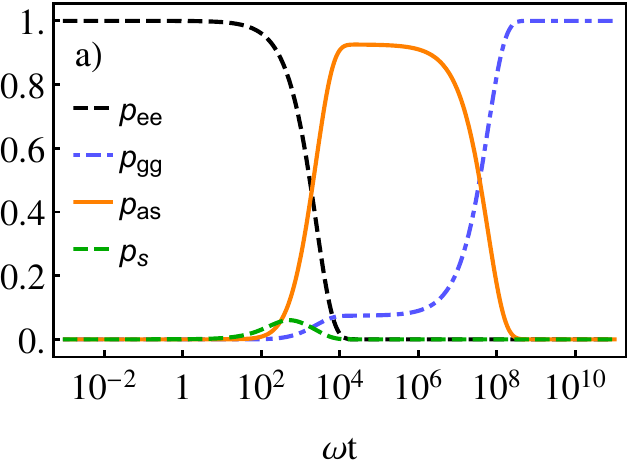}}
\end{minipage}
\hfill
\begin{minipage}[h]{0.49\linewidth}
\center{\includegraphics[width=0.72\linewidth]{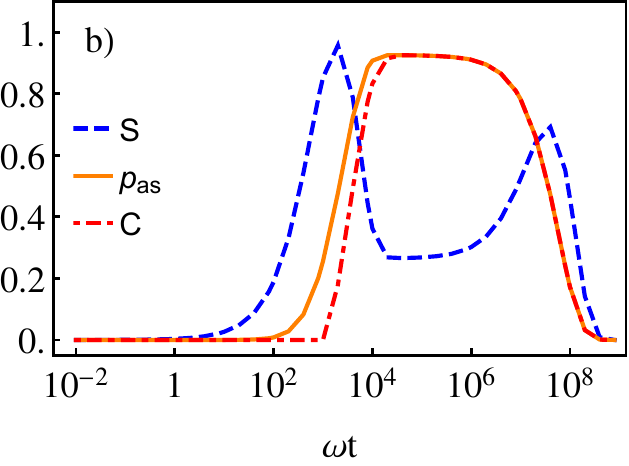} }
\end{minipage}
\caption{The dependencies of density matrix elements $p_{ee}$ (black dashed line), $p_{gg}$ (blue dash-dotted line), $p_{as}$ (orange solid line), $p_{s}$ (green dashed line) (a), and entropy (blue dashed line), concurrence (red dash-dotted line), antisymmetric state occupancy (orange solid line) (b) on time with one dephasing reservoir removed. }
\label{OneRes}
\end{figure*}

Numerical simulation shows (see Fig. \ref{OneRes}) that no significant changes of entropy, concurrence, or occupancy of the antisymmetric state  are observed, and time dynamics are similar to those from Fig. \ref{TwoRes}b. This is a consequence of the cross-relaxation processes described above.

\subsection{Disappearance of the entanglement in the case of absence of dephasing reservoirs}\label{B}

\begin{figure*}[ht]
\begin{minipage}[h]{0.49\linewidth}
\center{\includegraphics[width=0.72\linewidth]{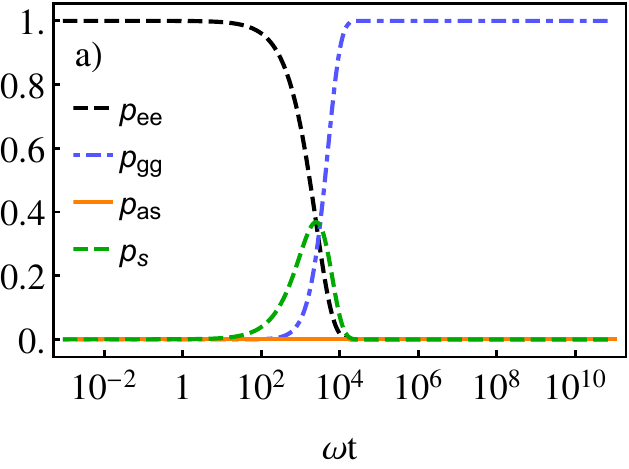}}
\end{minipage}
\hfill
\begin{minipage}[h]{0.49\linewidth}
\center{\includegraphics[width=0.72\linewidth]{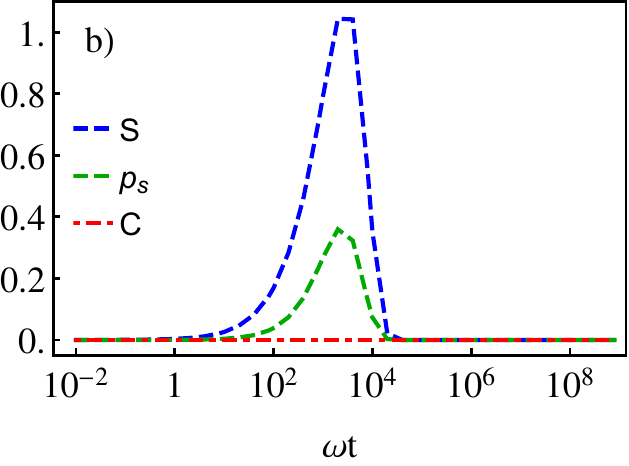} }
\end{minipage}
\caption{The dependencies of density matrix elements $p_{ee}$ (black dashed line), $p_{gg}$ (blue dash-dotted line), $p_{as}$ (orange solid line), $p_{s}$ (green dashed line) (a), and entropy (blue dashed line), concurrence (red dash-dotted line), symmetric state occupancy (green dashed line) (b) on time with both dephasing reservoirs removed.}
\label{ZeroRes}
\end{figure*}

In the second example, we remove both dephasing reservoirs ($\gamma_{\rm{1,dp}}=\gamma_{\rm{2,dp}}=0$), such that only the common radiative reservoir interacts with the TLSs. This case is analogous to the Dicke model of superradiance for two TLSs \cite{42}, such that only states $\left| {ee} \right\rangle$, $\left| s \right\rangle$, and $\left| {gg} \right\rangle$ are occupied (see Fig. \ref{ZeroRes}a). Relaxation of the system to the ground state with the rate  $\simeq{2\gamma _{\rm{rad}}}$ occurs. Concurrence, in all moments of time, is zero (Fig. \ref{ZeroRes}b, red dash-dotted line), which means that the system is in the disentangled state. Indeed, in the Dicke model of superradiance, entanglement is not generated \cite{43,44}. 

\subsection{Disappearance of entanglement in the local approach}\label{C}

In third example, we use local approach to describe the system. Under local approach, the Lindblad superoperators, arising from dephasing reservoirs, are assumed to coincide with the ones for noninteracting TLSs and have the form

\begin{equation}
\begin{array}{l}
{{\hat L}_{{\rm{rad}}{\rm{,1}}}} = {{\hat \sigma }_1} + {{\hat \sigma }_2} - \hat \sigma _1^\dag {{\hat \sigma }_1}{{\hat \sigma }_2} - {{\hat \sigma }_1}\hat \sigma _2^\dag {{\hat \sigma }_2},\ 
{{\hat L}_{{\rm{rad}}{\rm{,2}}}} = \hat \sigma _1^\dag {{\hat \sigma }_1}{{\hat \sigma }_2} + {{\hat \sigma }_1}\hat \sigma _2^\dag {{\hat \sigma }_2},\\
{{\hat L}_{{\rm{dp}}{\rm{1}}}} = {\hat \sigma _z^{\left( 1 \right)}},\ 
{{\hat L}_{{\rm{dp}}{\rm{2}}}} = {\hat \sigma _z^{\left( 2 \right)}}.
\end{array}
\end{equation}

\begin{figure*}
\begin{minipage}[h]{0.49\linewidth}
\center{\includegraphics[width=0.72\linewidth]{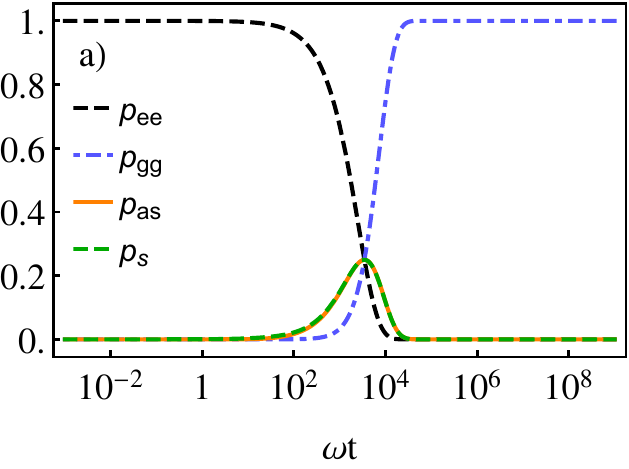}}
\end{minipage}
\hfill
\begin{minipage}[h]{0.49\linewidth}
\center{\includegraphics[width=0.72\linewidth]{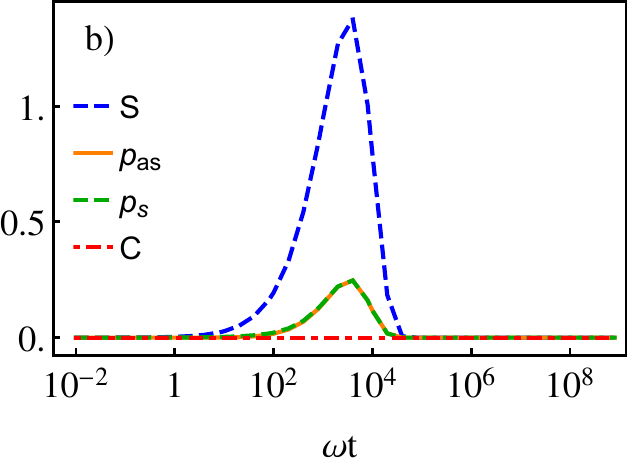} }
\end{minipage}
\caption{The dependencies of density matrix elements $p_{ee}$ (black dashed line), $p_{gg}$ (blue dash-dotted line), $p_{as}$ (orange solid line), $p_{s}$ (green dashed line) (a), and entropy (blue dashed line), concurrence (red dash-dotted line), antisymmetric state occupancy (orange solid line), symmetric state occupancy (green dashed line) (b) on time in local approach.}
\label{LocalAppr}
\end{figure*}

In the Fig. \ref{LocalAppr}, the results of this model are presented. Both symmetric and antisymmetric states have equal populations at all moments of time. Concurrence is zero in all moments of time, thus, entanglement does not occur.

\section{Analytical evaluation of the probability of subradiant state occupation and its lifetime}

To clarify the nature of the long-lived entangled state, from master equation (\ref{2}), we find the approximate  solution for $p_{as}$ in the limit $\hbar {\Omega} \gg kT$ for time scale $t \ge 2\gamma _{\rm{dp}}^{ - 1}{\rm{exp}}\left( {{\hbar \Omega}/{k T_{\rm{dp}}}} \right)$. We rewrite the master equation in terms of occupancies ${p_{ee}}$, ${p_s}$, ${p_{as}}$, ${p_{gg}}$ of the system eigenstates $\left| {ee} \right\rangle$, $\left| s \right\rangle$, $\left| {as} \right\rangle$, $\left| {gg} \right\rangle$ and obtain the following equations:
\begin{equation}
{\dot p_{ee}} =  - 2{\gamma _{{\rm{rad}}}}{p_{ee}},\ \ \ \ \ \ \ 
{\dot p_s} = 2{\gamma _{{\rm{rad}}}}{p_{ee}} - \left( {{\gamma _{{\rm{dp}}}}/2 + 2{\gamma _{{\rm{rad}}}}} \right){p_s} + \frac{{{\gamma _{{\rm{dp}}}}}}{2}{\rm{exp}}\left( { - 2\Omega /{T_{{\rm{dp}}}}} \right){p_{as}},
\label{an1}
\end{equation}
\begin{equation}
{\dot p_{as}} = \frac{{{\gamma _{{\rm{dp}}}}}}{2}{p_s} - \frac{{{\gamma _{{\rm{dp}}}}}}{2}{\rm{exp}}\left( { - 2\Omega /{T_{{\rm{dp}}}}} \right){p_{as}},\ \ \ \ \ \ \ 
{\dot p_{gg}} = 2{\gamma _{{\rm{rad}}}}{p_s},
\label{an2}
\end{equation}
with the initial condition
\begin{equation}
{p_{ee}}\left( 0 \right) = 1,\ \ \ \ \ \ \ {p_s}\left( 0 \right) = {p_{as}}\left( 0 \right) = {p_{gg}}\left( 0 \right) = 0.
\end{equation}

Because the case of strong coupling is examined, $\hbar \Omega  \gg kT$, at time scale $\gamma _{{\rm{rad}}}^{ - 1} \ll t \ll \gamma _{{\rm{dp}}}^{ - 1}{\rm{exp}}\left( {\hbar \Omega /k{T_{{\rm{dp}}}}} \right)$, we can suppose that the rate of transition from the sub- to superradiant state is zero. In such an assumption, we obtain the following quasi-stationary solution:
\begin{equation}
p_{ee}^{{\rm{qs}}} = p_s^{{\rm{qs}}} = 0,\ \ \ \ \ \ \ 
p_{as}^{{\rm{qs}}} = \frac{{{\gamma _{{\rm{dp}}}}/2}}{{2{\gamma _{{\rm{rad}}}} + {\gamma _{{\rm{dp}}}}/2}},\ \ \ \ \ \ \ 
p_{gg}^{{\rm{qs}}} = \frac{{2{\gamma _{{\rm{rad}}}}}}{{2{\gamma _{{\rm{rad}}}} + {\gamma _{{\rm{dp}}}}/2}}.
\label{StatSol}
\end{equation}
For the time scale $t \ge 2\gamma _{{\rm{dp}}}^{ - 1}{\rm{exp}}\left( {\hbar \Omega /k{T_{{\rm{dp}}}}} \right)$ we can use quasi-stationary solution (\ref{StatSol}) as the initial condition for Eqs. (\ref{an1}) -- (\ref{an2}). The answer takes the form:
\begin{equation}
p_{as}^{an}\left( t \right) \simeq \frac{{{\gamma _{\rm{dp}}/2}}}{{{2\gamma _{{\rm{rad}}}} + {\gamma _{\rm{dp}}/2}}}{\rm{exp}}\left( { - \frac{{{\gamma _{\rm{dp}}}{\gamma _{{\rm{rad}}}}t}}{{{\gamma _{\rm{dp}}/2} + {2\gamma _{{\rm{rad}}}}}}{\rm{exp}}\left( { - 2\hbar \Omega /k{T_{{\rm{dp}}}}} \right)} \right),\ \ \ \ \ \ \ 
p_{gg}^{an}\left( t \right) \simeq 1 - p_{as}^{an}\left( t \right).
\label{ApproxSol}
\end{equation}

\begin{figure}
    \centering
    \includegraphics[width=0.38\linewidth]{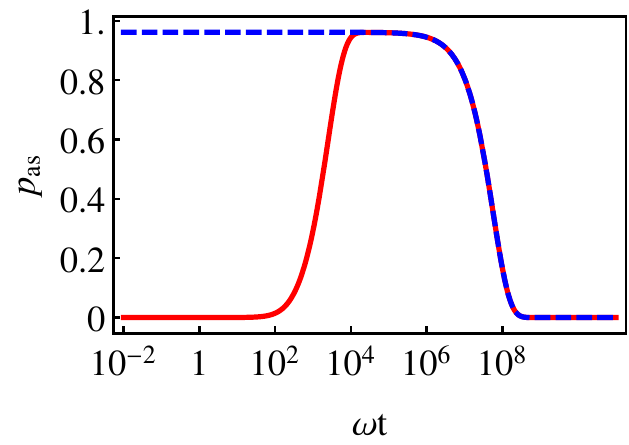}
    \caption{The dependence of the probability ${p_{as}}$ of the system to be in the antisymmetric state obtained from the numerical simulation of master equation (\ref{2}) (red line) and $p_{as}^{an}$ from analytical approximation (\ref{ApproxSol}) (blue dashed line).}
    \label{FigAnalit}
\end{figure}

Analytical evaluation of antisymmetric state occupancy, ${p^{an}_{as}}\left( t \right)$, is in a good agreement with exact solution, ${p_{as}}\left( t \right)$ at times $t \ge 2\gamma _{\rm{dp}}^{ - 1}{\rm{exp}}\left( {{\hbar \Omega}/{k T_{\rm{dp}}}} \right)$  (see Fig \ref{FigAnalit}). Eq. (\ref{ApproxSol}) shows that the characteristic lifetime of the subradiant state, in the limit  ${\gamma _{\rm{dp}}} \gg {\gamma _{{\rm{rad}}}}$, is of the order ${t_{{\rm{ent}}}} \simeq 0.5\gamma _{{\rm{rad}}}^{ - 1}{\rm{exp}}\left( {2{\hbar \Omega}/{k T_{\rm{dp}}}} \right) \gg 2\gamma _{\rm{dp}}^{ - 1},\,0.5\gamma _{{\rm{rad}}}^{ - 1}$, i.e., much greater than the characteristic time of dissipation and dephasing. 
Note that the long-lived state described above is a special case of metastable states proposed in \cite{45}. 

\section{Concurrence in the case of non-zero detuning between qubit frequencies}

\begin{figure}
    \centering
    \includegraphics[width=0.57\linewidth]{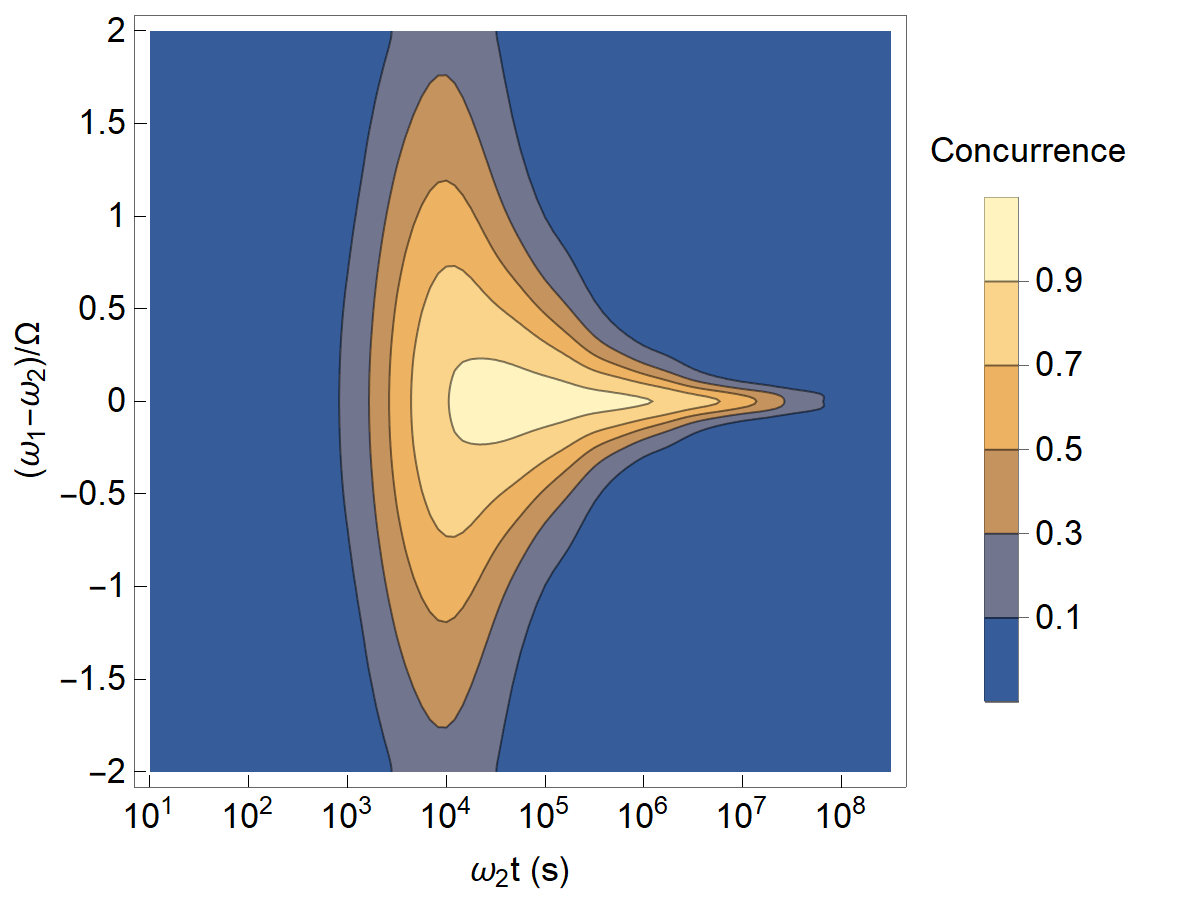}
    \caption{The dependence of concurrence on time $t$ and frequency detuning ${\omega _1} - {\omega _2}$. The frequency of the first qubit is changed while other parameters of the system are the same as in the main text.}
    \label{DitribConDetune}
\end{figure}

The long-lived entangled state is also exists in the case of non-zero detuning between qubit frequencies, $\omega_1 \ne \omega_2$. In the Appendix, we generalize the master equation for this case (see Eqs. (\ref{33}) -- (\ref{36})). Modeling this equation with initial condition $\hat \rho \left( 0 \right) = \left| {ee} \right\rangle \left\langle {ee} \right|$, one can get concurrence as a function of time and detuning (we fix  ${\omega _2}$ and change ${\omega _1}$) (see Fig \ref{DitribConDetune}). Its seen that the lifetime of the entangled state decreases when detuning grows. However, there exists detuning range ($\left| {{\omega _1} - {\omega _2}} \right| \sim 0.1\Omega $), for which the system stays in the entangled state (concurrence is grater than $0.9$) for the time which is much larger than the radiative and dephasing times. If detuning is out of this range, entangled state also is formed, but its lifetime is much shorter and its concurrence is smaller.

\section{Conclusion}

To conclude, in this work, we consider the system of two coupled two-level qubits with optical frequencies, which interact with the electromagnetic field of free space modes and with dephasing reservoirs. We show that when the coupling constant between qubits is much larger than the dephasing rate and dephasing rate is much larger than radiative rate, a long-lived mixed entangled state exists. The weight of the subradiant state in this mixed state, as well as concurrence, are equal to one up to the ratio ${\gamma _{\rm{rad}}}/{\gamma _{\rm{dp}}}$. The lifetime of this state is larger than the characteristic radiative time by the factor ${\rm{exp}}\left({2{\hbar \Omega}/{k T_{\rm{dp}}}} \right)$. 

The obtained result, in principle, can be extended to the case of many qubits. If one creates such interaction between qubits that symmetry with respect to permutation is protected at least approximately, then super- and subradiant states may still be eigenstates of the system. For example, when qubits are placed in the ring and approximate interaction Hamiltonian is $\hbar \Omega \sum\limits_{k = 1}^N {\left( {\hat \sigma _k^\dag {{\hat \sigma }_{k - 1}} + \hat \sigma _{k - 1}^\dag {{\hat \sigma }_k}} \right)}$ (where, by definition, ${\hat{\sigma} _0} = {\hat{\sigma} _n}$ and $N$ is even), there are the superradiant and subradiant states which for the case of one net excitation have the form  $\frac{1}{{\sqrt N }}\sum\limits_k {\left| {g...{e_k}...g} \right\rangle }$ and $\frac{1}{{\sqrt N }}\sum\limits_k {{{( - 1)}^k}\left| {g \ldots {e_k} \ldots g} \right\rangle }$, respectively (see \cite{27}). There is dephasing-induced transition from the super- to subradiant state, analogously to the case of two qubits. Careful investigation we leave on further works.

The obtained results pave the way for using dephasing as a resource for the creation of long-lived entanglement at experimentally realizable conditions. In the case of molecules, the characteristic dipole moment is $\simeq 1 {\ \rm{D}}$. If one achieves the characteristic distance between them $\simeq 10 {\ \rm{nm}}$, then the Rabi constant of interaction is ${\Omega}\simeq{10^{9}}{{\ \rm{s}}^{ - 1}}$. Thus, the factor ${\rm{exp}}\left( {2\hbar {\Omega}/k{T_{\rm{dp}}}} \right)$ is much larger than unity for temperatures  $T_{\rm{dp}} \le 5{\ \rm{mK}}$, and the lifetime of the entangled state can be $ \ge \gamma _{\rm{rad}}^{ - 1}\simeq{10^{ - 9}}\ \rm{s}$. If one considers semiconductor quantum dots with the characteristic dipole moment  $\simeq 50{\ \rm{D}}$ \cite{7,8} and distance between them $\simeq 10{\ \rm{nm}}$, one obtains  ${\Omega} \simeq 2 \cdot {10^{12}}{{\ \rm{s}}^{ - 1}}$ and the factor $\hbar {\Omega}/k{T_{\rm{dp}}}$ much larger than unity for the temperatures $T_{\rm{dp}} \le 10{\ \rm{K}}$. For example, for temperature $T_{\rm{dp}} \le {5\ \rm{K}}$ the factor $\exp\left({2\hbar {\Omega}/k{T_{\rm{dp}}}} \right) \simeq {10^3}$ and the lifetime of the entangled state may reach $ \simeq 1\ \mu {\rm{s}}$.

The predicted value of concurrence may be useful in contest of DLSZ protocol \cite{47,48}, where the value  $0.9 \pm 0.3$ of concurrence is used. In this work, we obtain the value $C \approx 0.9$. Also, this value is quite similar to maximal concurrence of two entangled photons achieved in recent experiments on optical quantum networks \cite{49}. 

Note that the simple model developed in this work may shed light on the recently demonstrated robustness of the subradiant state decay time for an ensemble of cold atoms with the increasing of temperature \cite{50}. In this regard, the toy-model that describes long-lived entanglement may be useful for qualitative analyses.

\section*{Appendix}
Here we derive the master equation (\ref{2}) and Lindblad superoperators (\ref{3}) from the main text and consider the case of non-zero detuning. 

To describe the interaction of two coupled qubits with the electromagnetic field of free space and separate dephasing reservoirs, we start with the Hamiltonian of the entire system. This Hamiltonian consists of three terms. ${\hat H_{\rm{R}}}$  represents the electromagnetic (EM) field of free space modes and degrees of freedom of dephasing reservoirs. The term ${\hat H_{\rm{S}}}$ describes the coupled qubits, and ${\hat H_{\rm{SR}}}$  is responsible for the interaction between the EM field and qubit dipole moments, and for the interaction between dephasing reservoirs and qubits. The corresponding Hamiltonians have the form:

\begin{equation}\hat H = {\hat H_{\rm{S}}} + {\hat H_{\rm{R}}} + {\hat H_{\rm{SR}}},\ 
{\hat H_{\rm{S}}} = \hbar {\omega _1}\hat \sigma _1^\dag {\hat \sigma _1} + \hbar {\omega _2}\hat \sigma _2^\dag {\hat \sigma _2} + \hbar \Omega \left( {\hat \sigma _1^\dag {{\hat \sigma }_2} + \hat \sigma _2^\dag {{\hat \sigma }_1}} \right),
\label{Hraw1}
\end{equation}
\begin{equation}
{\hat H_{\rm{R}}} = \sum\limits_{\rm{k}} {\hbar {\omega _{\rm{rad},{k}}}\hat b_{\rm{rad},{k}}^\dag {{\hat b}_{\rm{rad},{k}}}}  + \sum\limits_{\rm{k}} {\hbar {\omega _{\rm{1dp},{k}}}\hat b_{\rm{1dp},{k}}^\dag {{\hat b}_{\rm{1dp},{k}}}}  + \sum\limits_{\rm{k}} {\hbar {\omega _{\rm{2dp},{k}}}\hat b_{\rm{2dp},{k}}^\dag {{\hat b}_{\rm{2dp},{k}}}} , 
\end{equation}
\begin{equation}
{\hat H_{\rm{SR}}} = {\hat H_{\rm{S},\rm{1dp}}} + {\hat H_{\rm{S},\rm{2dp}}} + {\hat H_{\rm{S},\rm{rad}}},\ 
{\hat H_{\rm{S},\rm{rad}}} = \sum\limits_{\rm{k}} {\hbar \gamma _{\rm{k}}^{\rm{rad}}} \left( {\hat \sigma _1^\dag  + {{\hat \sigma }_1} + \hat \sigma _2^\dag  + {{\hat \sigma }_2}} \right)\left( {\hat b_{\rm{rad},{k}}^\dag  + {{\hat b}_{\rm{rad},{k}}}} \right), 
\end{equation}
\begin{equation}
{\hat H_{\rm{S},\rm{1dp}}} = \sum\limits_{\rm{k}} {\hbar \gamma _{\rm{k}}^{\rm{1dp}}\hat \sigma _1^\dag {{\hat \sigma }_1}} \left( {\hat b_{\rm{1dp},{k}}^\dag  + {{\hat b}_{\rm{1dp},{k}}}} \right),\ {\hat H_{\rm{S},\rm{2dp}}} = \sum\limits_{\rm{k}} {\hbar \gamma _{\rm{k}}^{\rm{2dp}}\hat \sigma _2^\dag {{\hat \sigma }_2}} \left( {\hat b_{\rm{2dp},{k}}^\dag  + {{\hat b}_{\rm{2dp},{k}}}} \right).
\end{equation}

Here ${\hat \sigma _1}$ and ${\hat \sigma _2}$ are lowering operators for the first and second qubit, respectively; ${\omega _1}$ and ${\omega _2}$ are transition frequencies. The last term in Eq. (\ref{Hraw1}) describes dipole-dipole interaction in the rotating wave approximation \cite{39}. The coupling constant between qubits is $\Omega  = \left( {{{\bf{d}}_1}{{\bf{d}}_2} - 3\left( {{{\bf{d}}_1}{\bf{n}}} \right)\left( {{{\bf{d}}_2}{\bf{n}}} \right)} \right)/\hbar {r^3}$, ${{\bf{d}}_{1,2}}$ are matrix elements of the first and second TLS dipole transitions, and ${\bf{n}}$ is the unit vector from one TLS to another. Operators ${\hat b_{\rm{1dp},k}},\;{\hat b_{\rm{2dp},k}}$ denote lowering operators for the k-th degree of freedom of the first and second dephasing reservoirs, respectively. We suppose that they obey commutation relations $\left[ {{{\hat b}_{\rm{1dp},k}},\;\hat b_{\rm{1dp},k'}^\dag } \right] = {\delta _{\rm{kk'}}}$, $\left[ {{{\hat b}_{\rm{2dp},k}},\;\hat b_{\rm{2dp},k'}^\dag } \right] = {\delta _{\rm{kk'}}}$,  and $\left[ {{{\hat b}_{\rm{1dp},k}},\;\hat b_{\rm{2dp},k'}^\dag } \right] = 0$; this means that the reservoirs are independent. Operator ${\hat b_{\rm{rad},k}}$ denotes the lowering operator for the k-th EM field mode of free space with commutation relation $\left[ {{{\hat b}_{\rm{rad},k}},\;\hat b_{\rm{rad},k'}^\dag } \right] = {\delta _{\rm{kk'}}}$ The constants $\gamma _{\rm{k}}^{\rm{1dp}}$, $\gamma _{\rm{k}}^{\rm{2dp}}$, and $\gamma _{\rm{k}}^{\rm{rad}}$ are interaction constants of the qubits with the k-th degree of freedom of dephasing and radiative reservoirs, respectively. 

To compute time dependence of the density matrix, first, we rewrite the von-Neuman equation in interaction representation using substitutions $\hat \rho (t) = U\hat{\tilde{\rho}} (t)U^\dag$, ${\hat H_{\rm{SR}}}(t) =U {\hat{\tilde{H}}_{\rm{SR}}}(t)U^\dag$, where $U=\exp \left( -i(\hat{H}_{\rm{S}}+\hat{H}_{\rm{R}})t/\hbar\right)$.

Note that all terms describing the system-reservoir interaction have the form $\hat S\hat R$, where $\hat S$ depends on system variables and $\hat R$ depends on the reservoir variables. Thus, each term in ${\hat H_{\rm{SR}}}$ can be presented in the form ${\hat{\tilde{H}}_{\rm{SR}}}(t) = \sum\limits_m {{{\hat{\tilde{S}}}_m}{{\hat{\tilde{R}}}_m}}$, (subindex m denotes each of reservoirs) where
$
{\hat{\tilde{S}}_m}(t) = \exp \left( {i\hat{H}_{\rm{S}}t/\hbar} \right){\hat S_m}(t)\exp \left( { - i\hat{H}_{\rm{S}}t/\hbar} \right)$, $
{\hat{\tilde{R}}_m}(t) = \exp \left( {i\hat{H}_{\rm{R}}t/\hbar} \right){\hat R_m}(t)\exp \left( { - i\hat{H}_{\rm{R}}t/\hbar} \right)
$. After this substitution we get the von-Neuman equation in the following form:

\begin{equation}
\frac{{\partial \hat{\tilde{\rho}} (t)}}{{\partial t}} = \frac{i}{\hbar }\left[ {\hat{\tilde{\rho}} ,{{\hat{\tilde{H}}}_{\rm{SR}}}(t)} \right].
\end{equation}

First, we consider $\omega_1=\omega_2$. To calculate ${\hat{\tilde{S}}_m}\left( t \right)$, one can establish the action of the operator $\exp \left( {i{{\hat H}_{\rm{S}}}t} \right)\hat \sigma (t)\exp \left( { - i{{\hat H}_{\rm{S}}}t} \right)$ on an arbitrary state $\left| \psi  \right\rangle$ and, then, write this action as a linear combination of operators from the basis set, which contains operators ${\hat \sigma _1}$, ${\hat \sigma _2}$, $\hat \sigma _1^\dag {\hat \sigma _2}$, ${\hat \sigma _1}{\hat \sigma _2}$, $\hat \sigma _1^\dag {\hat \sigma _1}{\hat \sigma _2}$, and $\hat \sigma _2^\dag {\hat \sigma _2}{\hat \sigma _1}$, their Hermitian conjugation, and Hermitian operators $\hat \sigma _1^\dag {\hat \sigma _1}$, $\hat \sigma _2^\dag {\hat \sigma _2}$, $\hat \sigma _1^\dag {\hat \sigma _1}\hat \sigma _2^\dag {\hat \sigma _2}$, $\hat 1$. After straightforward calculations, we obtain

\begin{equation}
\begin{array}{c}
{\hat{\tilde{\sigma}}_1}(t) + {\hat{\tilde{\sigma}}_2}(t) =\\ \left( {{{\hat \sigma }_1} + {{\hat \sigma }_2} - \hat \sigma _1^\dag {{\hat \sigma }_1}{{\hat \sigma }_2} - \hat \sigma _2^\dag {{\hat \sigma }_2}{{\hat \sigma }_1}} \right){e^{ - i\left( {\omega  + \Omega } \right)t}} + \left( {\hat \sigma _1^\dag {{\hat \sigma }_1}{{\hat \sigma }_2} + \hat \sigma _2^\dag {{\hat \sigma }_2}{{\hat \sigma }_1}} \right){e^{ - i\left( {\omega  - \Omega } \right)t}},
\end{array}
\end{equation}

\begin{equation}
\begin{array}{l}
\hat{\tilde{\sigma}}_1^\dag (t){{\hat{\tilde{\sigma}}}_1}(t)
 = \frac{1}{2}\left( {\hat \sigma _1^\dag {{\hat \sigma }_1} + \hat \sigma _2^\dag {{\hat \sigma }_2}} \right) + \frac{{\hat \sigma _1^\dag  + \hat \sigma _2^\dag }}{2} \cdot \frac{{{{\hat \sigma }_1} - {{\hat \sigma }_2}}}{2}{e^{i2\Omega t}} + \frac{{\hat \sigma _1^\dag  - \hat \sigma _2^\dag }}{2} \cdot \frac{{{{\hat \sigma }_1} + {{\hat \sigma }_2}}}{2}{e^{ - i2\Omega t}}.
\end{array}
\end{equation}
(Alternatively, one can use the Baker–Campbell–Hausdorff formula \cite{suzuki1985decomposition}.) The Lindblad superoperators can be obtained via standard procedure \cite{32,41,35}, assuming that the reservoirs are in thermal equilibrium. As a result, we obtain the following Lindblad equation

\begin{equation}
\begin{array}{l}
{{\dot{\hat{\rho}} }_{\rm{S}}}(t) =  - i{\hbar ^{ - 1}}[{{\hat H}_{\rm{S}}},{{\hat \rho }_{\rm{S}}}(t)] + \\
 \sum\limits_{\rm{k} = 1,2} {\sum\limits_{\rm{j} = 1,2,3} {\frac{{{\gamma _{{\rm{k}}{\rm{,dp}}}}\left( {2\Omega {\theta _{\rm{j}}}} \right)}}{2}\left( {2{{\hat L}_{{\rm{dp}}{\rm{,kj}}}}{{\hat \rho }_{\rm{S}}}(t)\hat L_{{\rm{dp}}{\rm{,kj}}}^\dag  - {{\hat \rho }_{\rm{S}}}(t)\hat L_{{\rm{dp}}{\rm{,kj}}}^\dag {{\hat L}_{{\rm{dp}}{\rm{,kj}}}} - \hat L_{{\rm{dp}}{\rm{,kj}}}^\dag {{\hat L}_{{\rm{dp}}{\rm{,kj}}}}{{\hat \rho }_{\rm{S}}}(t)} \right)} }  + \\
  \sum\limits_{\rm{k} = 1,2} {\frac{{{\gamma _{{\rm{rad}}}}\left( { - \left( {\omega  + {{( - 1)}^{{\rm{k}} - 1}}\Omega } \right)} \right)}}{2}\left( {2{{\hat L}_{{\rm{rad}}{\rm{,k}}}}{{\hat \rho }_{\rm{S}}}(t)\hat L_{{\rm{rad}}{\rm{,k}}}^\dag  - {{\hat \rho }_{\rm{S}}}(t)\hat L_{{\rm{rad}}{\rm{,k}}}^\dag {{\hat L}_{{\rm{rad}}{\rm{,k}}}} - \hat L_{{\rm{rad}}{\rm{,k}}}^\dag {{\hat L}_{{\rm{rad}}{\rm{,k}}}}{{\hat \rho }_{\rm{S}}}(t)} \right)}  + \\
  \sum\limits_{\rm{k} = 1,2} {\frac{{{\gamma _{{\rm{rad}}}}\left( {\left( {\omega  + {{( - 1)}^{{\rm{k}} - 1}}\Omega } \right)} \right)}}{2}\left( {2\hat L_{{\rm{rad}}{\rm{,k}}}^\dag {{\hat \rho }_{\rm{S}}}(t){{\hat L}_{{\rm{rad}}{\rm{,k}}}} - {{\hat \rho }_{\rm{S}}}(t){{\hat L}_{{\rm{rad}}{\rm{,k}}}}\hat L_{{\rm{rad}}{\rm{,k}}}^\dag  - {{\hat L}_{{\rm{rad}}{\rm{,k}}}}\hat L_{{\rm{rad}}{\rm{,k}}}^\dag {{\hat \rho }_{\rm{S}}}(t)} \right)} {\kern 1pt} ,
\end{array}
\label{LindBasic}
\end{equation}

where Lindblad superoperators have the form 

\begin{equation}
\begin{array}{l}
{{\hat L}_{{\rm{rad}}{\rm{,1}}}} = {{\hat \sigma }_1} + {{\hat \sigma }_2} - \hat \sigma _1^\dag {{\hat \sigma }_1}{{\hat \sigma }_2} - {{\hat \sigma }_1}\hat \sigma _2^\dag {{\hat \sigma }_2},\,\,{{\hat L}_{{\rm{rad}}{\rm{,2}}}} = \hat \sigma _1^\dag {{\hat \sigma }_1}{{\hat \sigma }_2} + {{\hat \sigma }_1}\hat \sigma _2^\dag {{\hat \sigma }_2},\\
{{\hat L}_{{\rm{dp}}{\rm{,11}}}} = \hat \sigma _1^\dag {{\hat \sigma }_1}/2 + \hat \sigma _2^\dag {{\hat \sigma }_2}/2,\,\,{{\hat L}_{{\rm{dp}}{\rm{,12}}}} = \left( {\hat \sigma _1^\dag  + \hat \sigma _2^\dag } \right)\left( {{{\hat \sigma }_1} - {{\hat \sigma }_2}} \right)/4,\,\,{{\hat L}_{{\rm{dp}}{\rm{,13}}}} = \left( {\hat \sigma _1^\dag  - \hat \sigma _2^\dag } \right)\left( {{{\hat \sigma }_1} + {{\hat \sigma }_2}} \right)/4,\\
{{\hat L}_{{\rm{dp}}{\rm{,21}}}} = \hat \sigma _1^\dag {{\hat \sigma }_1}/2 + \hat \sigma _2^\dag {{\hat \sigma }_2}/2,\,\,{{\hat L}_{{\rm{dp}}{\rm{,22}}}} = \left( {\hat \sigma _1^\dag  + \hat \sigma _2^\dag } \right)\left( {{{\hat \sigma }_2} - {{\hat \sigma }_1}} \right)/4,\,\,{{\hat L}_{{\rm{dp}}{\rm{,23}}}} = \left( {\hat \sigma _2^\dag  - \hat \sigma _1^\dag } \right)\left( {{{\hat \sigma }_1} + {{\hat \sigma }_2}} \right)/4,
\end{array}
\end{equation}
and parameter ${\theta _{\rm{j}}} = 0,\,\,1,\,\, - 1$ for ${\rm{j}} = 1,\,\,2,\,\,3$. The ${\gamma _m}\left( \omega  \right)$ functions are determined according to ${\gamma _m}\left( \omega  \right) = \int\limits_{ - \infty }^\infty  {d\tau \exp \left( { - i\omega \tau } \right)\left\langle {{{\hat{\tilde{R}}}_m}(t + \tau ){{\hat{\tilde{R}}}_m}(t)} \right\rangle }$, and, in the case of reservoirs which are in thermal equilibrium at temperature $T$, they satisfy the Kubo-Martin-Shwinger (KMS) condition ${\gamma _m}\left( \omega  \right) = \exp \left( { - \hbar \omega /kT} \right){\gamma _m}\left( { - \omega } \right)$.

If ${\omega _1} \ne {\omega _2}$ the states $\left| {ee} \right\rangle$ and $\left| {gg} \right\rangle$ remain the system eigenstates. Indeed, ${\hat H_{\rm{S}}}\left| {ee} \right\rangle  = \left( {{\omega _1} + {\omega _2}} \right)\left| {ee} \right\rangle$ and ${\hat H_{\rm{S}}}\left| {gg} \right\rangle  = 0\left| {gg} \right\rangle$. Other two eigenstates have the form 
\begin{equation}
\left|  +  \right\rangle  = \left( {\left| s \right\rangle  + \frac{\Delta }{{2\Omega  + \sqrt {{\Delta ^2} + 4{\Omega ^2}} }}\left| {as} \right\rangle } \right)/\sqrt {1 + {{\left( {\frac{\Delta }{{2\Omega  + \sqrt {{\Delta ^2} + 4{\Omega ^2}} }}} \right)}^2}} ,
\end{equation}
\begin{equation}
\left|  -  \right\rangle  = \left( {\left| {as} \right\rangle  + \frac{{2\Omega  - \sqrt {{\Delta ^2} + 4{\Omega ^2}} }}{\Delta }\left| s \right\rangle } \right)/\sqrt {1 + {{\left( {\frac{{2\Omega  - \sqrt {{\Delta ^2} + 4{\Omega ^2}} }}{\Delta }} \right)}^2}} ,
\end{equation}

with eigenvalues

\begin{equation}
{\omega _ + } = \frac{{\theta  + \sqrt {{\Delta ^2} + 4{\Omega ^2}} }}{2},\ \ \ \ \ \ \ \ 
{\omega _ - } = \frac{{\theta  - \sqrt {{\Delta ^2} + 4{\Omega ^2}} }}{2}
\end{equation}
respectively. Here $\theta  = {\omega _1} + {\omega _2}$, $\Delta  = {\omega _1} - {\omega _2}$.

To obtain new Lindblad equation, we, first, need to calculate ${\hat{\tilde{\sigma}}_{1,2}}(t) = \exp \left( {i{{\hat H}_{\rm{S}}}t} \right){\hat \sigma _{1,2}}(t)\exp \left( { - i{{\hat H}_{\rm{S}}}t} \right)$. After straightforward calculations, we obtain 

\begin{equation}
\begin{array}{*{20}{c}}
{{{\hat{\tilde{\sigma}}}_1}(t)}& = &{\frac{{{{\hat \sigma }_1}\frac{{y + 2\sqrt {{y^2}/4 + 1} }}{2} + {{\hat \sigma }_2}}}{{2\sqrt {{y^2}/4 + 1} }}\exp \left( {i\frac{{ - \theta  - \sqrt {{\Delta ^2} + 4{\Omega ^2}} }}{2}t} \right) - \frac{{\hat \sigma _1^\dag {{\hat \sigma }_1}{{\hat \sigma }_2}}}{{\sqrt {{y^2}/4 + 1} }}\exp \left( {i\frac{{ - \theta  - \sqrt {{\Delta ^2} + 4{\Omega ^2}} }}{2}t} \right)  }\\
{}&{}&{+\frac{{ - {{\hat \sigma }_1}\frac{{y - 2\sqrt {{y^2}/4 + 1} }}{2} - {{\hat \sigma }_2}}}{{2\sqrt {{y^2}/4 + 1} }}\exp \left( {i\frac{{ - \theta  + \sqrt {{\Delta ^2} + 4{\Omega ^2}} }}{2}t} \right) + \frac{{\hat \sigma _1^\dag {{\hat \sigma }_1}{{\hat \sigma }_2}}}{{\sqrt {{y^2}/4 + 1} }}\exp \left( {i\frac{{ - \theta  + \sqrt {{\Delta ^2} + 4{\Omega ^2}} }}{2}t} \right),}
\end{array}
\end{equation}
where $y = \left( {{\omega _1} - {\omega _2}} \right)/\Omega$. Analogous expression, up to replacement $1 \leftrightarrow 2$ and $y \to  - y$, is valid for ${\hat{\tilde{\sigma}}_2}(t)$. Using these expressions, we obtain 

\begin{equation}
\begin{array}{*{20}{c}}
{{{\hat{\tilde{\sigma}} }_1}(t) + {{\hat{\tilde{\sigma}} }_2}(t) + \hat{\tilde{\sigma}}_1^\dag (t) + \hat{\tilde{ \sigma}}_2^\dag (t) = }\\
{\frac{1}{{\sqrt {{y^2} + 4} }}\left[ {\left( {{{\hat \sigma }_1}\left( {1 + \frac{{y + \sqrt {{y^2} + 4} }}{2}} \right) + {{\hat \sigma }_2}\left( {1 + \frac{{ - y + \sqrt {{y^2} + 4} }}{2}} \right) - 2\hat \sigma _1^\dag {{\hat \sigma }_1}{{\hat \sigma }_2} - 2\hat \sigma _2^\dag {{\hat \sigma }_2}{{\hat \sigma }_1}} \right)\exp \left( {i\frac{{ - \theta  - \sqrt {{\Delta ^2} + 4{\Omega ^2}} }}{2}t} \right) + } \right.}\\
{ + \left( {{{\hat \sigma }_1}\left( { - 1 - \frac{{y - \sqrt {{y^2} + 4} }}{2}} \right) + {{\hat \sigma }_2}\left( { - 1 + \frac{{y + \sqrt {{y^2} + 4} }}{2}} \right) + 2\hat \sigma _1^\dag {{\hat \sigma }_1}{{\hat \sigma }_2} + 2\hat \sigma _2^\dag {{\hat \sigma }_2}{{\hat \sigma }_1}} \right)\exp \left( {i\frac{{ - \theta  + \sqrt {{\Delta ^2} + 4{\Omega ^2}} }}{2}t} \right) + }\\
{\left. { + h.\;c.} \right],}
\end{array}
\end{equation}

\begin{equation}
\begin{array}{*{20}{c}}
{\hat{\tilde{\sigma}}_1^\dag (t){{\hat{\tilde{\sigma}}}_1}(t) = }
{\frac{1}{{{y^2} + 4}}\left[ {2\left( {\hat \sigma _1^\dag {{\hat \sigma }_1}\frac{{{y^2} + 2}}{2} + \hat \sigma _2^\dag {{\hat \sigma }_2}} \right) + y\left( {\hat \sigma _1^\dag {{\hat \sigma }_2} + \hat \sigma _2^\dag {{\hat \sigma }_1}} \right) } \right.}\\
{-\left( {\hat \sigma _1^\dag \frac{{y + \sqrt {{y^2} + 4} }}{2} + \hat \sigma _2^\dag } \right)\left( {{{\hat \sigma }_1}\frac{{y - \sqrt {{y^2} + 4} }}{2} + {{\hat \sigma }_2}} \right)\exp \left( {i\sqrt {{\Delta ^2} + 4{\Omega ^2}} t} \right) }\\
{-\left. {\left( {\hat \sigma _1^\dag \frac{{y - \sqrt {{y^2} + 4} }}{2} + \hat \sigma _2^\dag } \right)\left( {{{\hat \sigma }_1}\frac{{y + \sqrt {{y^2} + 4} }}{2} + {{\hat \sigma }_2}} \right)\exp \left( { - i\sqrt {{\Delta ^2} + 4{\Omega ^2}} t} \right)} \right],}
\end{array}
\end{equation}

\begin{equation}
\begin{array}{*{20}{c}}
{\hat{\tilde{\sigma}}_2^\dag (t){{\hat{\tilde{\sigma}}}_2}(t) = }
{\frac{1}{{{y^2} + 4}}\left[ {2\left( {\hat \sigma _2^\dag {{\hat \sigma }_2}\frac{{{y^2} + 2}}{2} + \hat \sigma _1^\dag {{\hat \sigma }_1}} \right) - y\left( {\hat \sigma _1^\dag {{\hat \sigma }_2} + \hat \sigma _2^\dag {{\hat \sigma }_1}} \right) } \right.}\\
{-\left( {\hat \sigma _2^\dag \frac{{ - y + \sqrt {{y^2} + 4} }}{2} + \hat \sigma _1^\dag } \right)\left( {{{\hat \sigma }_2}\frac{{ - y - \sqrt {{y^2} + 4} }}{2} + {{\hat \sigma }_1}} \right)\exp \left( {i\sqrt {{\Delta ^2} + 4{\Omega ^2}} t} \right)}\\
{-\left. {\left( {\hat \sigma _2^\dag \frac{{ - y - \sqrt {{y^2} + 4} }}{2} + \hat \sigma _1^\dag } \right)\left( {{{\hat \sigma }_2}\frac{{ - y + \sqrt {{y^2} + 4} }}{2} + {{\hat \sigma }_1}} \right)\exp \left( { - i\sqrt {{\Delta ^2} + 4{\Omega ^2}} t} \right)} \right].}
\end{array}
\end{equation}

Using standard procedure for derivation of Linblad superoperators, one can get

\begin{equation}
\begin{array}{l}
{{\dot{\hat{\rho}} }_{\rm{S}}}(t) =  - i{\hbar ^{ - 1}}[{{\hat H}_{\rm{S}}},{{\hat \rho }_{\rm{S}}}(t)] + \\
 + \sum\limits_{{\rm{k}} = 1,2} {\sum\limits_{{\rm{j}} = 1,2,3} {\frac{{{\gamma _{{\rm{k}}{\rm{,dp}}}}\left( {\sqrt {{\Delta ^2} + 4{\Omega ^2}} {\theta _{\rm{j}}}} \right)}}{2}\left( {2{{\hat L}_{{\rm{dp}}{\rm{,kj}}}}{{\hat \rho }_{\rm{S}}}(t)\hat L_{{\rm{dp}}{\rm{,kj}}}^\dag  - {{\hat \rho }_{\rm{S}}}(t)\hat L_{{\rm{dp}}{\rm{,kj}}}^\dag {{\hat L}_{{\rm{dp}}{\rm{,kj}}}} - \hat L_{{\rm{dp}}{\rm{,kj}}}^\dag {{\hat L}_{{\rm{dp}}{\rm{,kj}}}}{{\hat \rho }_{\rm{S}}}(t)} \right)} }  + \\
 + \sum\limits_{{\rm{k}} = 1,2} {\frac{{{\gamma _{{\rm{rad}}}}\left( { - \left( {\theta  + {{( - 1)}^{{\rm{k}} - 1}}\sqrt {{\Delta ^2} + 4{\Omega ^2}} } \right)/2} \right)}}{2}\left( {2{{\hat L}_{{\rm{rad}}{\rm{,k}}}}{{\hat \rho }_{\rm{S}}}(t)\hat L_{{\rm{rad}}{\rm{,k}}}^\dag  - {{\hat \rho }_{\rm{S}}}(t)\hat L_{{\rm{rad}}{\rm{,k}}}^\dag {{\hat L}_{{\rm{rad}}{\rm{,k}}}} - \hat L_{{\rm{rad}}{\rm{,k}}}^\dag {{\hat L}_{{\rm{rad}}{\rm{,k}}}}{{\hat \rho }_{\rm{S}}}(t)} \right)}  + \\
 + \sum\limits_{{\rm{k}} = 1,2} {\frac{{{\gamma _{{\rm{rad}}}}\left( {\left( {\theta  + {{( - 1)}^{{\rm{k}} - 1}}\sqrt {{\Delta ^2} + 4{\Omega ^2}} } \right)/2} \right)}}{2}\left( {2\hat L_{{\rm{rad}}{\rm{,k}}}^\dag {{\hat \rho }_{\rm{S}}}(t){{\hat L}_{{\rm{rad}}{\rm{,k}}}} - {{\hat \rho }_{\rm{S}}}(t){{\hat L}_{{\rm{rad}}{\rm{,k}}}}\hat L_{{\rm{rad}}{\rm{,k}}}^\dag  - {{\hat L}_{{\rm{rad}}{\rm{,k}}}}\hat L_{{\rm{rad}}{\rm{,k}}}^\dag {{\hat \rho }_{\rm{S}}}(t)} \right)} {\kern 1pt} ,
\end{array}
\label{33}
\end{equation}

where Lindblad superoperators have the form 

\begin{equation}
\begin{array}{l}
{{\hat L}_{{\rm{rad}}{\rm{,1}}}} = \frac{1}{{\sqrt {{y^2} + 4} }}\left( {{{\hat \sigma }_1}\left( {1 + \frac{{y + \sqrt {{y^2} + 4} }}{2}} \right) + {{\hat \sigma }_2}\left( {1 + \frac{{ - y + \sqrt {{y^2} + 4} }}{2}} \right) - 2\hat \sigma _1^\dag {{\hat \sigma }_1}{{\hat \sigma }_2} - 2\hat \sigma _2^\dag {{\hat \sigma }_2}{{\hat \sigma }_1}} \right),\,\,\\
{{\hat L}_{{\rm{rad}}{\rm{,2}}}} = \frac{1}{{\sqrt {{y^2} + 4} }}\left( {{{\hat \sigma }_1}\left( { - 1 - \frac{{y - \sqrt {{y^2} + 4} }}{2}} \right) + {{\hat \sigma }_2}\left( { - 1 + \frac{{y + \sqrt {{y^2} + 4} }}{2}} \right) + 2\hat \sigma _1^\dag {{\hat \sigma }_1}{{\hat \sigma }_2} + 2\hat \sigma _2^\dag {{\hat \sigma }_2}{{\hat \sigma }_1}} \right),
\end{array}
\label{34}
\end{equation}

\begin{equation}
\begin{array}{l}
{{\hat L}_{{\rm{dp}}{\rm{,11}}}} = \frac{1}{{{y^2} + 4}}\left( {2\left( {\hat \sigma _1^\dag {{\hat \sigma }_1}\frac{{{y^2} + 2}}{2} + \hat \sigma _2^\dag {{\hat \sigma }_2}} \right) + y\left( {\hat \sigma _1^\dag {{\hat \sigma }_2} + \hat \sigma _2^\dag {{\hat \sigma }_1}} \right)} \right),\,\\
\,{{\hat L}_{{\rm{dp}}{\rm{,12}}}} =  - \frac{1}{{{y^2} + 4}}\left( {\hat \sigma _1^\dag \frac{{y + \sqrt {{y^2} + 4} }}{2} + \hat \sigma _2^\dag } \right)\left( {{{\hat \sigma }_1}\frac{{y - \sqrt {{y^2} + 4} }}{2} + {{\hat \sigma }_2}} \right),\,\,\\
{{\hat L}_{{\rm{dp}}{\rm{,13}}}} =  - \frac{1}{{{y^2} + 4}}\left( {\hat \sigma _1^\dag \frac{{y - \sqrt {{y^2} + 4} }}{2} + \hat \sigma _2^\dag } \right)\left( {{{\hat \sigma }_1}\frac{{y + \sqrt {{y^2} + 4} }}{2} + {{\hat \sigma }_2}} \right),
\end{array}
\label{35}
\end{equation}

\begin{equation}
\begin{array}{l}
{{\hat L}_{{\rm{dp}}{\rm{,21}}}} = \frac{1}{{{y^2} + 4}}\left( {2\left( {\hat \sigma _2^\dag {{\hat \sigma }_2}\frac{{{y^2} + 2}}{2} + \hat \sigma _1^\dag {{\hat \sigma }_1}} \right) - y\left( {\hat \sigma _1^\dag {{\hat \sigma }_2} + \hat \sigma _2^\dag {{\hat \sigma }_1}} \right)} \right),\,\\
\,{{\hat L}_{{\rm{dp}}{\rm{,22}}}} =  - \frac{1}{{{y^2} + 4}}\left( {\hat \sigma _2^\dag \frac{{ - y + \sqrt {{y^2} + 4} }}{2} + \hat \sigma _1^\dag } \right)\left( {{{\hat \sigma }_2}\frac{{ - y - \sqrt {{y^2} + 4} }}{2} + {{\hat \sigma }_1}} \right),\,\,\\
{{\hat L}_{{\rm{dp}}{\rm{,23}}}} =  - \frac{1}{{{y^2} + 4}}\left( {\hat \sigma _2^\dag \frac{{ - y - \sqrt {{y^2} + 4} }}{2} + \hat \sigma _1^\dag } \right)\left( {{{\hat \sigma }_2}\frac{{ - y + \sqrt {{y^2} + 4} }}{2} + {{\hat \sigma }_1}} \right).
\end{array}
\label{36}
\end{equation}

\section*{Funding}
Russian Science Foundation (20-72-10057). 

\section*{Acknowledgments}
The study was supported by a grant from Russian Science Foundation (project No. 20-72-10057). E.S.A. thank foundation for the advancement of theoretical physics and mathematics “Basis”.

\bibliography{refs}

\end{document}